\documentclass[runningheads]{llncs}

\usepackage[title]{appendix}
\usepackage{comment}
\usepackage{amsmath}
\usepackage{amsfonts}
\usepackage{amssymb}
\usepackage{graphicx}
\usepackage{algorithm}
\usepackage{algorithmic}

\usepackage{xspace}
\usepackage{booktabs}
\usepackage[inline]{enumitem}
\newenvironment{enumin}[3][.]{\begin{enumerate*}[label=\arabic*),itemjoin={{#2 }},itemjoin*={{#2 #3 }},after={#1}]}{\end{enumerate*}}
\usepackage{tikz}
\usetikzlibrary{positioning,calc}
\tikzset{loc/.style={draw,rectangle,rounded corners,inner sep=0.5mm},
         t/.style={->,>=stealth,shorten >= 0.5mm,shorten <= 0.5mm}}
\usepackage{hyperref}

\newcommand{\figref}[1]{Fig.~\ref{#1}}

\newcommand{\R}{\ensuremath{\mathbb{R}}\xspace}
\newcommand{\Rnn}{\ensuremath{\mathbb{R}_{\geq 0}}\xspace}
\newcommand{\eps}{\ensuremath{\varepsilon}\xspace}
\newcommand{\norm}[1]{\ensuremath{\Vert #1 \Vert}\xspace}
\newcommand{\rest}[2]{\ensuremath{#1\!\!\downharpoonright_{#2}}}
\newcommand{\dom}{\texttt{dom}\xspace}

\newcommand{\F}{\ensuremath{\mathcal{S}}\xspace}

\newcommand{\x}{\ensuremath{\mathbf{x}}\xspace}
\newcommand{\y}{\ensuremath{\mathbf{y}}\xspace}
\newcommand{\va}{\ensuremath{\mathbf{a}}\xspace}
\newcommand{\m}{\ensuremath{\mathbf{m}}\xspace}
\newcommand{\p}{\ensuremath{\mathbf{p}}\xspace}
\newcommand{\q}{\ensuremath{\mathbf{q}}\xspace}
\newcommand{\vb}{\ensuremath{\mathbf{b}}\xspace}
\newcommand{\origin}{\ensuremath{\mathbf{0}}\xspace}
\newcommand{\X}{\ensuremath{\mathcal{X}}\xspace}
\newcommand{\B}{\ensuremath{\mathcal{B}_\eps}\xspace}
\newcommand{\PP}{\ensuremath{P}\xspace}
\newcommand{\cpoly}{\ensuremath{\mathbb{P}}\xspace}

\renewcommand{\H}{\ensuremath{\mathcal{H}}\xspace}
\newcommand{\Loc}{\ensuremath{\textit{Loc}}\xspace}
\newcommand{\E}{\ensuremath{\textit{E}}\xspace}
\newcommand{\Flow}{\ensuremath{\textit{Flow}}\xspace}
\newcommand{\I}{\ensuremath{\textit{Inv}}\xspace}
\newcommand{\Grd}{\ensuremath{\textit{Grd}}\xspace}
\newcommand{\loc}{\ensuremath{\ell}\xspace}
\newcommand{\dwell}[1]{\ensuremath{\xrightarrow{#1}}}
\newcommand{\jumplabel}{\ensuremath{\mathit{jmp}}\xspace}
\newcommand{\jump}{\ensuremath{\xrightarrow{\jumplabel}}}
\newcommand{\exc}{\ensuremath{\sigma_\pi}\xspace}
\newcommand{\Hd}{\ensuremath{\mathcal{H}_d}\xspace}

\newcommand{\qedex}{\,\hfill\ensuremath{\triangleleft}}
\newcommand{\Con}{red!30!white}
\newcommand{\Coff}{blue!20!white}

\newcommand{\pwl}{PWL\xspace}
\newcommand{\lha}{LHA\xspace}

\newcommand{\map}{\ensuremath{\mathcal{M}}\xspace}

\begin{document}

\title{Synthesis of Parametric Hybrid Automata \\ from Time Series}

\author{
Miriam García Soto\inst{1}
\orcidID{0000-0003-2936-5719}
\and \\
Thomas A. Henzinger\inst{2}
\orcidID{0000-0002-2985-7724}
\and \\
Christian Schilling\inst{3}
\orcidID{0000-0003-3658-1065}
}
\authorrunning{M. Garc\'ia Soto et al.}

\institute{
Complutense University of Madrid, Spain \\
\email{miriamgs@ucm.es}
\and
IST Austria, Klosterneuburg, Austria \\
\email{tah@ist.ac.at}
\and
Aalborg University, Aalborg, Denmark \\
\email{christianms@cs.aau.dk}
}

\maketitle

\begin{abstract}
	We propose an algorithmic approach for synthesizing linear hybrid 
	automata from time-series data.
	Unlike existing approaches, our approach provides a whole family of models with the same discrete structure but different dynamics.
	Each model in the family is guaranteed to capture the input data up to a precision error $\varepsilon$, in the following sense:
	For each time series, the model contains an execution that is $\varepsilon$-close to the data points.
	Our construction allows to effectively choose a model from this family with minimal precision error $\varepsilon$.
	We demonstrate the algorithm's efficiency and its ability to find precise models in two case studies.

\end{abstract}

\section{Introduction}

Mathematical models are ubiquitous across all sciences~\cite{FriggH20}, from systems biology~\cite{KlippLWK16} to epidemiology~\cite{VynnyckyW10} to cyber-physical systems~\cite{LeeS17}.
The construction of such models is a central challenge in science~\cite{Silvert01}.
One main benefit of \emph{mathematical} models is the clearly defined semantics, which make these models amenable to automatic analysis (such as simulation~\cite{Fishwick07,Raimondi08} and verification~\cite{ClarkeHVB18,Platzer18,AlthoffFG20}).
Another main benefit that is usually desired is interpretability for high-level reasoning.

Hybrid automata~\cite{AlurCHH92,Henzinger00} are a prominent class of interpretable models with mixed continuous and discrete behavior.
They are particularly suitable in biological domains~\cite{SinghaniaSJT11,LiuB20}, where systems typically evolve continuously but are subject to internal and external events, and in cyber-physical domains~\cite{KhaitanM15}, where physical entities interact with digital devices.
In a nutshell, the evolution of a hybrid automaton follows a differential equation associated with one of several locations (or modes), until a discrete event leads to a different location.

\smallskip

In this paper we address the problem of synthesizing a linear hybrid automaton (LHA)~\cite{AlurCHH92} from a set of time series.
The informal goal of model synthesis is that the model \emph{captures} the data well.
What it means to ``capture well'' is difficult to formalize.
Here we adopt the recent notion of \emph{\eps-capturing} from García et al.~\cite{SotoHSZ19,SotoHS21}, which requires that, for each time series in the input data, the LHA must expose an execution that stays \eps-close to all data points (see \figref{fig:thermostat_synthesized} for an illustration).
In~\cite{SotoHSZ19,SotoHS21}, the value of \eps is fixed in the problem input.
Here we consider \eps a parameter, which we associate with
a \emph{family of parametric models}: LHA whose continuous dynamics are not fixed yet.
Each possible fixation of the continuous dynamics corresponds to an instantiated LHA.
All instantiated LHA associated with a concrete value of \eps have the property that they \eps-capture the data.
We can then effectively search for an \eps-capturing LHA with the minimal value of \eps, whose behavior intuitively best resembles the data.

Our algorithm consists of two phases.
In the first phase we synthesize the discrete structure of the LHA by fixing the set of locations and mapping data points in the time series to the locations.
We propose an algorithm to obtain this mapping based on clustering.
In the second phase we construct the parameter space, which is a polyhedron that associates \eps to all possible instantiated LHA (i.e., fixations of continuous dynamics) that \eps-capture the data.
We select a concrete LHA by minimizing the value of \eps, for which we solve a linear program.

We evaluate the algorithm in two case studies.
In the first case study we investigate the scalability in terms of the different input parameters; we can synthesize a seven-dimensional model with $15$ locations from $12{,}000$ data points in $15$ minutes, which shows that the algorithm is applicable in practice.
In the second case study we use the algorithm to synthesize a model for a biological system (regulation of a cell cycle) in less than half a minute.

\paragraph{Related work.}

Synthesizing models is known to different communities as \emph{system identification}, \emph{process mining}, or \emph{model learning}.
Models that are akin to hybrid automata have been studied extensively
in control theory; while the main aim in control theory is to find a
controller for a system,
which is outside the scope of the present paper, there is still a large
body of works on system identification~\cite{PaolettiJFV07,GarulliPV12}.
Many of these approaches focus on input-output models, such as
autoregressive exogenous (ARX) models and in particular the switched
(SARX)~\cite{HashambhoyV05,Ozay16} and piecewise
(PWARX)~\cite{FerrariM03,RollBL04,NakadaTK05,JuloskiWH05,BemporadGPV05}
versions, and focus on single-input single-output (SISO) systems, but
there are also works on multiple-input multiple-output (MIMO)
systems~\cite{HuangWM04,VerdultV04,BakoV08}.
SARX and PWARX models can be seen as restricted linear hybrid automata where the locations form a state-space partition and the switching behavior is deterministic. This allows to reduce the synthesis problem to a parameter-optimization problem.
The second phase of our algorithm also uses a reduction to parameter optimization, but the parameter space is different and our model class is more general.

In computer science, several approaches learn hybrid automata from input-output traces or time series.
Similar to our approach, the works in~\cite{MedhatRBF15,BartocciDGMNQ20} first use clustering to learn the discrete structure, but they employ different techniques, such as Angluin's algorithm for learning a finite automaton, and do not provide minimality guarantees for the result.
Other approaches construct automata whose discrete structure is acyclic~\cite{NiggemannSVMB12} respectively cyclic~\cite{GrosuMYERS07}, or a deterministic model with urgent transitions~\cite{LamraniBG18}.
The work in~\cite{SummervilleOM17} exhaustively constructs all possible models for optimizing a cost function, while in our approach the enumeration is only symbolic and we choose a model by solving a linear program, which scales favorably.
A recent work shows that timed automata can be effectively learned from traces with a genetic algorithm~\cite{TapplerALL19}; learning timed automata has orthogonal challenges: they form a subclass of LHA where all variables are clocks with constant rate $1$ and hence no continuous dynamics need to be learned, but the discrete dynamics are more complex than in this work.
The work in~\cite{YangX22} provides a framework for identifying deterministic models with affine dynamics from input-output traces, while we identify nondeterministic models.
Our works in~\cite{SotoHSZ19,SotoHS21} proposed the notion of \eps-capturing that we adopt here; those works synthesize a model from single traces online, but the algorithms are not scalable for offline usage of realistic dimension and size.

\paragraph{Outline.}

In Section~\ref{sec:terminology} we fix the terminology.
In Section~\ref{sec:problem} we formalize the synthesis problem and describe our solution on a high level.
The low-level descriptions of the two phases of the algorithm follow in Section~\ref{sec:phase1} and Section~\ref{sec:phase2}.
We evaluate the algorithm in Section~\ref{sec:evaluation} and conclude in Section~\ref{sec:conclusion}.

\section{Terminology}\label{sec:terminology}

\paragraph{Euclidean sets.}
We write $\x$ for points $(x_1 , \ldots , x_n)$ in $\R^n$ and consider the infinity norm $\norm{\x} = \max_{x_i} |x_i|$.
The \emph{ball} of radius $\eps \in \Rnn$ around a point $\x \in \R^n$ is $\B(\x) = \{\y \in \R^n : \norm{\x - \y} \leq \eps\}$.
The \emph{$\eps$-bloating} of $\X \subseteq \R^n$ is $\X \oplus \B(\origin) = \{\x + \y : \x \in \X, \norm{\y} \leq \eps\}$.
A \emph{polyhedron} over $\R^n$ is a finite intersection of \emph{constraints} $\va^T \x \leq b$ where $\va \in \R^n$ and $b \in \R$.
Let $\cpoly_n$ be the set of all $n$-dimensional polyhedra.
An \emph{interval} is written $[a, b] = \{x : a \leq x \leq b\} \subseteq \R$.

\paragraph{Functions.}
Given a function $f: A \to B$, let $\dom(f) \subseteq A$ denote its 
domain.
Let $\rest{f}{D}$ denote the restriction of $f$ to set $D \subseteq \dom(f)$.
A continuous function $f : [0, T] \to \R^n$ is a 
\emph{piecewise-linear} (\emph{\pwl}) \emph{function} with $k$ 
pieces if there exists a triple $(I, M, \x_0)$ where $I$ is a $k$-tuple of
consecutive time intervals $[t_0, t_1], [t_1, t_2], \ldots , [t_{k-1}, t_k]$ with $[0, T] = \bigcup_{1\leq i \leq k} [t_{i-1}, t_i]$, $M$ is a $k$-tuple
of slope vectors $\m_i \in \R^n$, and $\x_0 \in \R^n$ is the initial
state $f(t_0) = \x_0$, such that each $\rest{f}{[t_{i-1}, t_i]}$ is a
solution of the differential equation $\dot{\x}(t) = \m_i$, for all $i = 1,
\ldots, k$.
We refer to the line segments $\rest{f}{[t_{i-1}, t_i]}$ as the \emph{pieces} of $f$.
A \emph{time-series} $s : D \to \R^n$ maps time points $t$ from a 
finite set $D \subseteq \Rnn$ to data points $s(t)$.
There is a one-to-one correspondence between \pwl functions and time series:
A \pwl function $f$ over $I = ([t_0, t_1], [t_1, t_2], \ldots , [t_{k-1}, 
t_k])$ induces a time series as the restriction $s = \rest{f}{D}$ to time 
points $D = \{t_0, t_1, \ldots, t_k\}$, and $s$ induces $f$ as the 
piecewise-linear interpolation of the data points.
Thus we may refer to, e.g., the pieces of a time series.
The \emph{distance} between a \pwl function $f$ and a time series $s$ with
$\dom(\rest{f}{\dom(s)}) = \dom(s)$ is $d(f, s) = \max_{t \in \dom(s)}
\norm{f(t) - s(t)}$.

\paragraph{Linear hybrid automata.}
An $n$-dimensional \emph{linear hybrid automaton} (\lha) 
\cite{AlurCHH92,Henzinger00} is a tuple $\H = (\Loc, \E, \Flow, \I, 
\Grd)$, where
\begin{enumin}{,}{and}
	\item
	$\Loc$ is the finite set of locations
	\item
	$\E \subseteq \Loc \times \Loc$ is the transition relation
	\item
	$\Flow: \Loc \to \R^n$ is the flow function
	\item
	$\I: \Loc \to \cpoly_n$ is the invariant function
	\item
	$\Grd: \E \to \cpoly_n$ is the guard function
\end{enumin}
Our \lha model does not have assignments along the 
transitions and is also called switched linear system~\cite{Liberzon03}.
We also consider partially defined hybrid automata without flows, 
invariants, or guards assigned.
This \emph{discrete structure} $\H_d = (\Loc, \E)$ only consists of locations and transitions.

The semantics of \lha are described by the set of executions.
A \emph{state} of an \lha is a pair $(\loc, \x)$ of a location $\loc \in \Loc$
and a point $\x \in \I(\loc)$ in the invariant.
An \emph{execution} $\sigma$ of an \lha evolves continuously 
according to 
the flow function in each location.
The execution starts in some state $(\loc_1, \x_1)$ and the continuous evolution follows the 
constant differential equation $\dot{\x} = \Flow(\loc_1)$ while 
satisfying the invariant $\I(\loc_1)$ for some dwell time $\delta \in \Rnn$. The 
execution can instantaneously switch 
locations, from a state $(\loc_1, \x_2)$ to another state $(\loc_2, 
\x_2)$, if there is a transition $(\loc_1, \loc_2) \in \E$ and the guard 
$\Grd(\loc_1, \loc_2)$ contains $\x_2$.
The projection of an execution 
$\sigma$ to the second component is a \pwl function, which we denote 
by $\exc$.
We use the following compact notation for executions, where $\delta_i 
\in \Rnn$ (for $i \geq 1$) denotes the duration of a dwell action and 
\jumplabel denotes a switch:
$
\sigma \equiv (\loc_1, \x_1)
\dwell{\delta_1} (\loc_1, \x_2)
\jump (\loc_2, \x_2)
\dwell{\delta_2} (\loc_2, \x_3)
\jump (\loc_3, \x_3) \ \cdots
$

\section{Synthesis of \eps-close linear hybrid automata}\label{sec:problem}

In this section we formalize the synthesis problem that we address in this paper and give a high-level overview of our approach to solve it.
Given a time series, we want to construct an LHA that captures the data up to a given precision.
We first formalize the notion of capturing.

\begin{definition}[\eps-capturing~\cite{SotoHSZ19}]
	Given a time series $s$ and a value $\eps \in \Rnn$, we say that
	an \lha $\H$ \emph{\eps-captures} $s$ if there exists an execution
	$\sigma$ of $\H$ such that $d(\exc, s) \leq \eps$.
	We also say that $s$ and $\exc$ (resp.\ $s$ and $\sigma$) are \emph{\eps-close}.
\end{definition}

Our goal is to construct an \lha that \eps-captures several time series.

\begin{problem}[\eps-close synthesis~\cite{SotoHSZ19}]
	Given a finite set of time series $\F$ and a value $\eps \in \Rnn$,
	construct an \lha \H that \eps-captures each $s$ in $\F$.
\end{problem}

As we observed in~\cite{SotoHSZ19}, it is straightforward to find a solution to the problem even for $\eps = 0$ by simply introducing a fresh location for each piece of the time series.
Such a model does not aggregate nor generalize the information in the data and is hence of little use.
To obtain a reasonable model, one needs to add another bound to the problem, e.g., by fixing the discrete structure.

We address this observation in a two-phase algorithm.
In the first phase we fix the discrete structure $\Hd$ of the LHA, where we try to reuse the locations for multiple time series (or pieces therein).
In the second phase we instantiate the model for the smallest possible value of \eps under the given discrete structure.
Thus in this paper we consider a synthesis problem where we do not fix the value of \eps and rather find a sufficiently small value for \eps automatically.

\begin{problem}[\eps-minimal synthesis]\label{prob:synthesis_minimal}
	Given a finite set of time series $\F$ and a discrete structure $\Hd$,
	find the minimal value $\eps \in \Rnn$ and an instantiation \H of $\Hd$ such that \H \eps-captures each $s$ in $\F$.
\end{problem}

\subsection{Synthesis algorithm}

In the next two sections we describe our algorithm to solve the above synthesis problem, but first we give a high-level overview of the algorithm.
Our algorithm computes a parametric family of LHA that all \eps-capture the given data.
The LHA share the same discrete structure but differ in the continuous dynamics.
Since \eps itself is a parameter of that construction, we can then choose an LHA with a minimal value for \eps (which is not necessarily unique) from that family.

Our goal is that the final LHA has an \eps-close execution for each time series.
To simplify the theoretical presentation, we will use the following conceptual view on our algorithm.
Instead of synthesizing an LHA directly, we synthesize \eps-close 
executions.
These executions then induce an LHA.

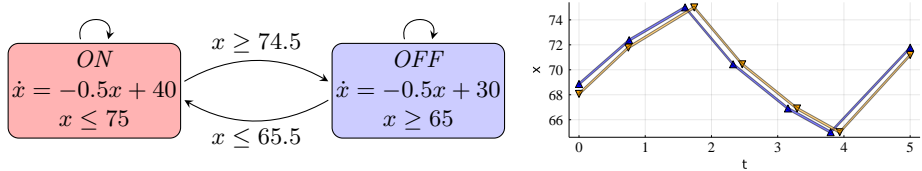
\begin{figure}[tb]
	\begin{minipage}{.55\textwidth}
		\begin{tikzpicture}
	\node[loc,fill=\Con] (on) {\begin{tabular}{@{} c @{}}\textit{ON} \\ $\dot{x} = -0.5 x + 40$ \\ $x \leq 75$\end{tabular}};
	\node[loc,fill=\Coff,right=20mm of on] (off) {\begin{tabular}{@{} c @{}}\textit{OFF} \\ $\dot{x} = -0.5 x + 30$ \\ $x \geq 65$\end{tabular}};
	\draw[t,bend left] ($(on.east) + (0,1mm)$) to node[above] {$x \geq 74.5$} ($(off.west) + (0,1mm)$);
	\draw[t,loop above,looseness=3] (on) to (on);
	\draw[t,bend left] ($(off.west) + (0,-1mm)$) to node[below] {$x \leq 65.5$} ($(on.east) + (0,-1mm)$);
	\draw[t,loop above,looseness=3] (off) to (off);
\end{tikzpicture}
	\end{minipage}
	\hfill
	\begin{minipage}{.43\textwidth}
		\includegraphics[width=\textwidth,keepaspectratio]{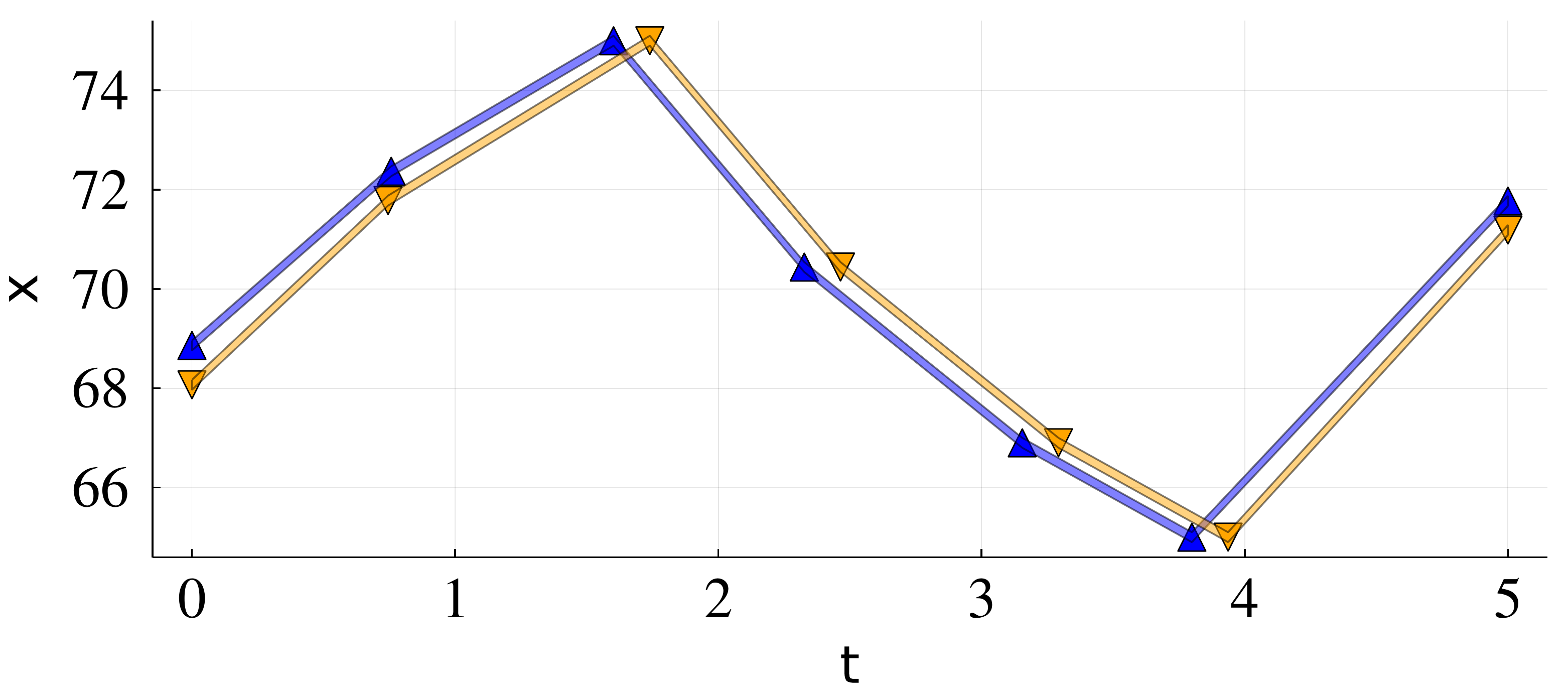}
	\end{minipage}
	\caption{Left: A hybrid automaton.
	Right: Two time series (triangle markers) obtained from sampling two executions of the automaton, and induced \pwl functions.}
	\label{fig:thermostat_original}
\end{figure}

As mentioned, our algorithm proceeds in two phases.
In the first phase we fix the discrete structure of the executions (and thus of the resulting LHA).
In the second phase we construct the space of continuous dynamics to be assigned to the locations, depending on the value \eps.
For \lha, this space is a polyhedron, which we call the 
\emph{flow polyhedron}.
We then choose concrete continuous dynamics from the flow polyhedron to instantiate concrete executions (and thus an LHA).
We explain each step of the algorithm using the following running example.

\begin{example}[running example]
	We consider two time series in one dimension:
	\vspace*{-2mm}
	\begin{align*}
		t_1 &= & (0.00, && 0.76, && 1.59, && 2.32, && 3.15, && 3.79, && 
5.00) \\[-1mm]
		d_1 &= & (68.91, && 72.41, && 75.00, && 70.44, && 66.90, && 
65.00, && 71.81) \\
		t_2 &= & (0.0, && 0.75, && 1.61, && 2.33, && 3.16, && 3.76, && 
5.00) \\[-1mm]
		d_2 &= & (68.16, && 71.85, && 74.70, && 70.22, && 66.75, && 
65.00, && 71.92) \\[-7mm]
	\end{align*}
	We obtained the time series from two random trajectories of a hybrid automaton modeling a simple thermostat controller, all given in \figref{fig:thermostat_original}.
	Note that the original continuous dynamics are
	described by an affine differential equation, which cannot be
	expressed with an \lha.
	(We round all numbers to two digits, which explains small inconsistencies over the course of this running example.)
	\qedex
\end{example}

\section{Synthesis algorithm, Phase 1: Discrete structure}\label{sec:phase1}

In this section we describe Phase~$1$ of the synthesis algorithm.
The input is a finite set of time series.
The output is a mapping from each piece of the time series (resp.\ the induced \pwl functions) to a \emph{symbolic location} (i.e., a location label).
Together with the order of the pieces in the time series, as we explain below, this mapping already fixes the discrete structure $\Hd$ of the LHA.

\subsection{Simplification of the time series}\label{sec:preprocessing}

In the first step of our algorithm, we preprocess the time series by removing some data points for better stability of the second step (we explain this connection later).
Note that for Phase~$2$ we again use the original time series, so correctness is not affected.
The goal is to merge consecutive pieces in the time series with similar slopes, i.e., such that the linear interpolation is a good approximation.
In our implementation we use a variant of the Ramer-Douglas-Peucker algorithm~\cite{Ramer72,DouglasP73} where we consider time as another dimension.
We shortly recall this algorithm but refer to the literature for details.
Following a divide-and-conquer scheme, the algorithm starts with only the first and last point of the time series, connects them with a line segment, finds the point \x with the largest distance from the line segment, and, unless this distance is small enough, repeats the process recursively for the corresponding two parts before and after \x.

\subsection{Assignment of symbolic locations}\label{sec:clustering}

The goal of the first phase is to determine the discrete structure $\Hd$ of the resulting LHA.
For each time series with $p$ pieces we synthesize a 
corresponding \emph{symbolic execution} of the prospective LHA.
These are executions that do not yet contain information about the continuous state, but the discrete state is already determined, i.e., we fix the sequence of visited locations $\loc_1, \dots, \loc_p$ together with the points in time when the execution switches to a new location.
(Here we restrict ourselves to switching in synchrony with the time series.)
Thus each symbolic execution consists of a (timed) sequence of symbolic locations.
It is easy to see that, by ignoring time, these sequences induce the discrete structure $\Hd$ of an LHA: the set of locations is the union of all locations occurring in the sequences, and there is a transition for each consecutive pair of locations.
Formally, for a symbolic execution associated with a time series with $p$ pieces, the discrete structure $\H_d = (\Loc, \E)$ is given by $\Loc = \{\loc_1, \dots, \loc_p\}$ and $ \E = \{ (\loc_i, \loc_{i+1}) : i = 1, \dots, p-1 \}$, and the generalization to sets of symbolic executions consists of the union of these locations and transitions.

\smallskip

Given a time series with $p$ pieces and a set of symbolic locations $\{ 
\loc_1, \dots, \loc_\lambda \}$, a symbolic execution as described 
above is merely a mapping from the pieces to location labels, which we call
$\map : \{1, \dots, p\} \to \{1, \dots, \lambda\}$.
Our algorithm is parametric in the concrete way to obtain this mapping.
Typically we are interested in finding an LHA with a small number of locations.
Thus the implicit requirement for the mapping is to share locations for multiple pieces.

In our implementation we obtain the mapping using a variant of the $k$-means clustering algorithm~\cite{Lloyd82}.
The input to the clustering algorithm are the slopes of the \pwl 
functions induced by the time series.
The $k$-means algorithm requires to specify upfront the number of clusters $k$, which corresponds to the number of locations in our setting.
If the intended number of locations is already known in advance, this algorithm can be used directly.
Otherwise, to find a good value of $k$ automatically, we use a common refinement loop by starting with some value for $k$ (e.g., $k = 1$) and then increasing $k$ until the clustering error (which is defined as the sum of the squared Euclidean distance of each point to its associated cluster center) does not decrease substantially anymore.

\begin{algorithm}[t]
	\caption{Assignment of a symbolic location to each piece of a set of time series.
	Line~\ref{line:preprocessing} is optional and can be implemented with the identity.
	Line~\ref{line:clustering} can be implemented with $k$-means, which can also provide a good value for $\lambda$ ($= k$) if not specified in the input (as described in Sect.~\ref{sec:clustering}).}
	\label{alg:preprocessing}
	\begin{algorithmic}[1]
		\REQUIRE A set of time series $\F = \{s_1, \dots, s_r\}$ and optionally a number of locations $\lambda$
		\ENSURE A mapping from the pieces to symbolic locations and a number of locations
		\STATE $\F'$ := simplify($\F$) \COMMENT{see Sect.~\ref{sec:preprocessing}} \label{line:preprocessing}
		\STATE $\map, \lambda$ := assign$\_$location$\_$labels$\_$to$\_$pieces($\F', \lambda$) \COMMENT{see Sect.~\ref{sec:clustering}} \label{line:clustering}
		\RETURN{$\map, \lambda$}
	\end{algorithmic}
\end{algorithm}

The $k$-means algorithm is sensitive to the initial choice of the cluster centers.
The preprocessing step proposed in Sect.~\ref{sec:preprocessing} increases the stability in this regard.
As initial candidates for the cluster centers we choose the first $k$ slopes induced by the simplified time series.
This choice results in candidates that are sufficiently different in 
practice and thus $k$-means yields more robust clusters.

We summarize the main steps of Phase~$1$ in Algorithm~\ref{alg:preprocessing}.

\begin{example}[cont'd]
	The input to the clustering algorithm are the slope values of the two time series.
	In the table below we list the clustering cost for different numbers of clusters $k$, together with the relative improvement compared to $k-1$:
	\begin{center}
		\begin{tabular}{c @{\quad} c @{\quad} c @{\quad} c @{\quad} c @{\quad} c @{\quad} c @{\quad} c @{\quad} c}
			\toprule
			clusters ($k$) & 1 & 2 & 3 & 4 & 5 & 6 & 7 & 8 \\
			\midrule
			cost & 259.76 & 17.07 & 11.80 & 2.46 & 0.78 & 0.09 & 0.04 & 0.01 \\
			\midrule
			rel.\ [\%] & -- & 0.93 & 0.31 & 0.79 & 0.68 & 0.89 & 0.60 & 0.61 \\
			\bottomrule
		\end{tabular}
	\end{center}

	The table suggests that good values for $k$ are $2$, $4$, or $6$.
	To obtain a small model, here we settle for $k = 2$ locations.
	The associated (one-dimensional) cluster centers (representing slopes) are $4.53$ and $-4.46$.
	For both time series, the assigned clusters are $(1, 1, 2, 2, 2, 1)$, corresponding to the symbolic location $\loc_1$ for the pieces $1$, $2$, $6$ and symbolic location $\loc_2$ for the other three pieces.
	\qedex
\end{example}

\section{Synthesis algorithm, Phase 2: Continuous dynamics}\label{sec:phase2}

In this section we describe Phase~$2$ of the synthesis algorithm.
The input is a finite set of time series together with a discrete structure $\Hd$ obtained in Phase~$1$, which is represented by the mapping $\map$ assigning a symbolic location to each piece of the time series.
The output is an \lha $\H$ and a value for \eps such that \H \eps-captures the time series.
As mentioned before, we describe how to obtain an \eps-close \emph{corresponding execution} for each time series.

\subsection{Construction of the flow polyhedron}\label{sec:polyhedron}

In the first step, we construct the flow polyhedron \PP, which represents the set of all possible continuous dynamics such that the corresponding executions are \eps-close to the time series.
Here \eps itself is a dimension of \PP.
For technical reasons, we construct a new flow polyhedron for each time series.

Assume that we have $n$-dimensional data in the form of $r$ time 
series and we want to synthesize an \lha with $\lambda$ locations.
Say that we consider a time series with $p$ pieces.
Then \PP is a polyhedron with $\lambda n + r n + 1$ dimensions.
The first $\lambda n$ dimensions represent the location slopes.
The next $r n$ dimensions represent the coordinates of the initial states $\x_0^{(j)}$ of the $j$-th execution.
(These $\x_0^{(j)}$ are auxiliary dimensions which we are not interested in.)
The last dimension is \eps.

Next we describe the constraints of \PP.
These constraints express that the distance between the time series and the execution is less than \eps (and thus the execution \eps-captures the time series).
We need to express the symbolic value of the execution, $\x_k$, at each time point $t_k$ of the time series.
Let $\q_k$ be the $k$-th data point of the time series, starting at $k = 0$.
For each data point we have $2 n$ constraints (i.e., $2 n (p + 1)$ constraints in total) to express the requirement $\x_k \in \B(\q_k)$.
In $n = 1$ dimension, for each $k$ we express the requirement with the two constraints $x_k - \eps \leq q_k$ and $x_k + \eps \geq q_k$.
In $n > 1$ dimensions we have such constraints in each dimension.

It remains to explain how to express the term $x_k$.
For $k = 0$ we represent $\x_0$ with the dedicated variables $x_0^{(\cdot)}$.
For $k > 0$ we rewrite $x_k$ using the following identity: $x_k = x_0 + \sum_{j=1}^k (t_j - t_{j-1}) m^{(j)}$.
The time points $t_j$ are known constants and the $m^{(j)}$ are the slope variables for the $j$-th piece (recall that we have associated the pieces with locations in advance).

Below we formalize the flow polyhedron for $r = 1$ time series.

\begin{definition}
	Given a time series $s$ with $p$ pieces and an associated mapping
	$\map : \{1, \dots, p\} \to \{1, \dots, \lambda\}$, the \emph{flow 
	polyhedron} $\PP_s$ is defined as
	$$
	\begin{aligned}
	\{ (\m_1, \ldots, \m_\lambda, 
	\x_0, \eps) \in \R^{\lambda n + n} \times \Rnn \mid 
	\x_0 &\in \B(s(t_0)),\\
	\x_0 + (t_1 - t_0) \m_{\map(1)} &\in \B(s(t_1)), \\
	\x_0 + (t_1 - t_0) \m_{\map(1)} + (t_2 - t_1) \m_{\map(2)} 
	&\in \B(s(t_2)),\\
	&~\vdots \\
	\x_0 + (t_1 - t_0) \m_{\map(1)} + \ldots +
	(t_p - t_{p-1}) \m_{\map(p)} &\in \B(s(t_p))  \}.
	\end{aligned}
	$$
\end{definition}

\begin{example}[cont'd]
	Our example has $n = 1$ dimension, $\lambda = 2$ locations, and $r = 2$ time series.
	The flow polyhedron consists of five variables $(m_1, m_2, x_0^{(1)}, x_0^{(2)}, \eps)$.
	Here $m_1$ and $m_2$ represent the slopes of the two locations, $x_0^{(1)}$ and $x_0^{(2)}$ represent the initial state of the first resp.\ second execution, and \eps represents the allowed distance between the time series and the executions.
	Below we show the $14$ constraints for the first execution:
	\begin{center}
	$
	\scriptstyle
	\begin{matrix}
	& & & & x_0^{(1)} & - & \eps &\leq& 68.91 \\
	0.76 m_1 & & & + & x_0^{(1)} & - & \eps &\leq& 72.41 \\
	1.59 m_1 & & & + & x_0^{(1)} & - & \eps &\leq& 75.00 \\
	1.59 m_1 & + & 0.72 m_2 & + & x_0^{(1)} & - & \eps &\leq& 70.44\\
	1.59 m_1 & + & 1.55 m_2 & + & x_0^{(1)} & - & \eps &\leq& 66.90 \\
	1.59 m_1 & + & 2.20 m_2 & + & x_0^{(1)} & - & \eps &\leq& 65.00 \\
	2.80 m_1 & + & 2.20 m_2 & + & x_0^{(1)} & - & \eps &\leq& 71.81
	\end{matrix}
	\hfill \vrule \hfill
	\begin{matrix}
	& & & - & x_0^{(1)} & - & \eps &\leq& -68.91 \phantom{\qedex} \\
	- 0.76 m_1 & & & - & x_0^{(1)} & - & \eps &\leq& -72.41 
	\phantom{\qedex} \\
	- 1.59 m_1 & & & - & x_0^{(1)} & - & \eps &\leq& -75.00 
	\phantom{\qedex} \\
	- 1.59 m_1 & - & 0.72 m_2 & - & x_0^{(1)} & - & \eps &\leq& -70.44 
	\phantom{\qedex} \\
	- 1.59 m_1 & - & 1.55 m_2 & - & x_0^{(1)} & - & \eps &\leq& -66.90 
	\phantom{\qedex} \\
	- 1.59 m_1 & - & 2.20 m_2 & - & x_0^{(1)} & - & \eps &\leq& -65.00 
	\phantom{\qedex} \\
	- 2.80 m_1 & - & 2.20 m_2 & - & x_0^{(1)} & - & \eps &\leq& -71.81 
	\qedex
	\end{matrix}
	$
\end{center}
\end{example}

Note that, for multiple time series, each flow polyhedron only constrains $n$ dimensions of the $r n$ dimensions reserved for the initial states $x_0^{(\cdot)}$.
The need for the separate dimensions will become clear when we aggregate the different flow polyhedra in the next step.
Any feasible point inside the polyhedron \PP represents a concrete 
execution in an \lha that \eps-captures the time series.
We formalize this statement after defining the corresponding \lha in 
the next step.

\subsection{The common solution space}\label{sec:intersect}

In the first phase we implicitly fixed the discrete evolution of the 
executions, which also induced the discrete structure of the \lha we 
want to synthesize.
In the previous step we obtained the flow polyhedra $\PP_s$, one for each time series $s$.
In the next steps we combine these results to obtain concrete executions by assigning the continuous states.
The concrete executions also induce the final \lha, i.e., we 
assign continuous dynamics, invariants, and guards.

Since we want to obtain one \lha to \eps-capture \emph{all} time 
series, we need to find compatible values for the dynamics and \eps.
For that purpose we can just intersect all flow polyhedra.
Let $\PP_\H = \bigcap_{s \in \F} \PP_s$ be the polyhedron resulting from this intersection.
Note that, since we used disjoint dimensions for the 
$\x_0^{(\cdot)}$ for different executions, the initial states are not 
shared in $\PP_\H$.
(We note that intersecting polyhedra in constraint representation is a constant-time 
operation.)

\subsection{Choice of minimizing parameters}\label{sec:minimize}

Now we have to choose \emph{any} feasible point $\p$ in 
$\PP_\H$.
We argue that the most interesting points are those that minimize \eps, since they correspond to executions that are closest to the original data.
(In applications where further constraints should be considered, other choices are possible.)
Minimizing a polyhedron in the dimension of \eps means to solve the corresponding linear program with objective function \eps, which is efficient in practice.
We remark that $\PP_\H$ is bounded in the dimension of \eps from 
below by $0$, so this minimization always returns a proper solution 
$\p =  (\m_1, \ldots, \m_\lambda, \x_0^{(1)}, \ldots, 
\x_0^{(r)}, \eps)$.
The point $\p$ contains a number for each dimension.
The first $\lambda n$ numbers are the slope values for the locations, 
in the order they have been specified.
The next $r n$ numbers are the values of $\x_0^{(\cdot)}$ for the 
different executions (note again that we do not need these numbers).
The last number is the corresponding value for \eps.

\subsection{Construction of the final LHA}\label{sec:lha_construction}

Next we describe, for a given time series $s_i$ over time instants $t_0, 
t_1, \ldots, t_p$, the execution that is induced by the above point 
$\p$.
Let $\m_1, \dots, \m_\lambda$ be the slopes taken from the point and
$\map$ be the mapping from the pieces of $s_i$ to the associated
location (e.g., $\loc_{\map(1)}$ is the location for the first piece, with slope
$\m_{\map(1)}$) obtained in Algorithm~\ref{alg:preprocessing}.
The execution is a \pwl function whose pieces have the same duration as the pieces of $s_i$.
As defined before, the execution starts at $\x_0 = \x_0^{(i)}$ and the end point of the $k$-th piece is $\x_k = \x_0 + \sum_{j=1}^p (t_j - t_{j-1}) \m_{\map(j)}$.
\begin{align*}
	(\loc_{\map(1)}, \x_0) &\dwell{t_1 - t_0} 
(\loc_{\map(1)}, \x_0 + (t_1 - t_0) \m_{\map(1)}) \\
	&\jump (\loc_{\map(2)}, \x_0 + (t_1 - t_0) 
\m_{\map(1)}) \\
	&\dwell{t_2 - t_1} (\loc_{\map(2)}, \x_0 + (t_1 - t_0) 
\m_{\map(1)} + (t_2 - t_1) \m_{\map(2)}) \\[-2mm]
	&\hspace*{5mm} \vdots \\[-2mm]
	&\dwell{t_p - t_{p-1}} (\loc_{\map(p)}, \x_0 + \sum_{j=1}^p 
(t_j - t_{j-1}) \m_{\map(j)})
\end{align*}

\begin{algorithm}[t]
	\caption{Synthesis algorithm.}
	\label{alg:synthesis}
	\begin{algorithmic}[1]
		\REQUIRE A set of time series $\F = \{s_1, \dots, s_r\}$, a number of locations $\lambda$, and a mapping from the pieces of each time series to symbolic locations $\map$
		\ENSURE An \lha \H and a minimal value \eps such that \H \eps-captures 
		all elements of \F
		\FOR {$s \in \F$}
			\STATE $\PP_s$ := flow$\_$polyhedron($s, \map, \lambda$) \COMMENT{see Sect.~\ref{sec:polyhedron}}
		\ENDFOR
		\STATE $\PP_\H$ := $\bigcap_{s \in \F}^r \PP_s$ \COMMENT{see Sect.~\ref{sec:intersect}}
		\STATE $\mathit{slopes}$, $\eps$ := choose$\_$minimizing$\_$point($\PP_\H$) \COMMENT{see Sect.~\ref{sec:minimize}} \label{line:minimize}
		\STATE $\H$ := construct$\_$automaton($\F, \map, \mathit{slopes}, \eps$) \COMMENT{see Sect.~\ref{sec:lha_construction}}
		\RETURN{\H, \eps}
	\end{algorithmic}
\end{algorithm}

We have not yet described the invariants and guards of the resulting 
\lha.
We say that a data point in the time series is associated with a location 
if the preceding or the succeeding piece in the time series is assigned 
that location in the mapping from Algorithm~\ref{alg:preprocessing}.
Similarly, a data point is associated with the transition $(\loc_i, \loc_j)$ if the preceding piece is associated with $\loc_i$ and the succeeding piece is associated with $\loc_j$.
A sufficient condition for our construction to be correct is:
define the invariant of each location as the \eps-bloated convex hull around all data points associated with it, and define the guard of each transition as the \eps-bloated union around all data points associated with it.
In our implementation we use the \eps-bloated interval hull in both cases.
That is, we take the smallest box around all data points as defined above and then extend the box in each direction by \eps.
We summarize the main steps of Phase~$2$ in Algorithm~\ref{alg:synthesis}.

\begin{figure}[tb]
	\centering
	\includegraphics[width=.49\textwidth,keepaspectratio]{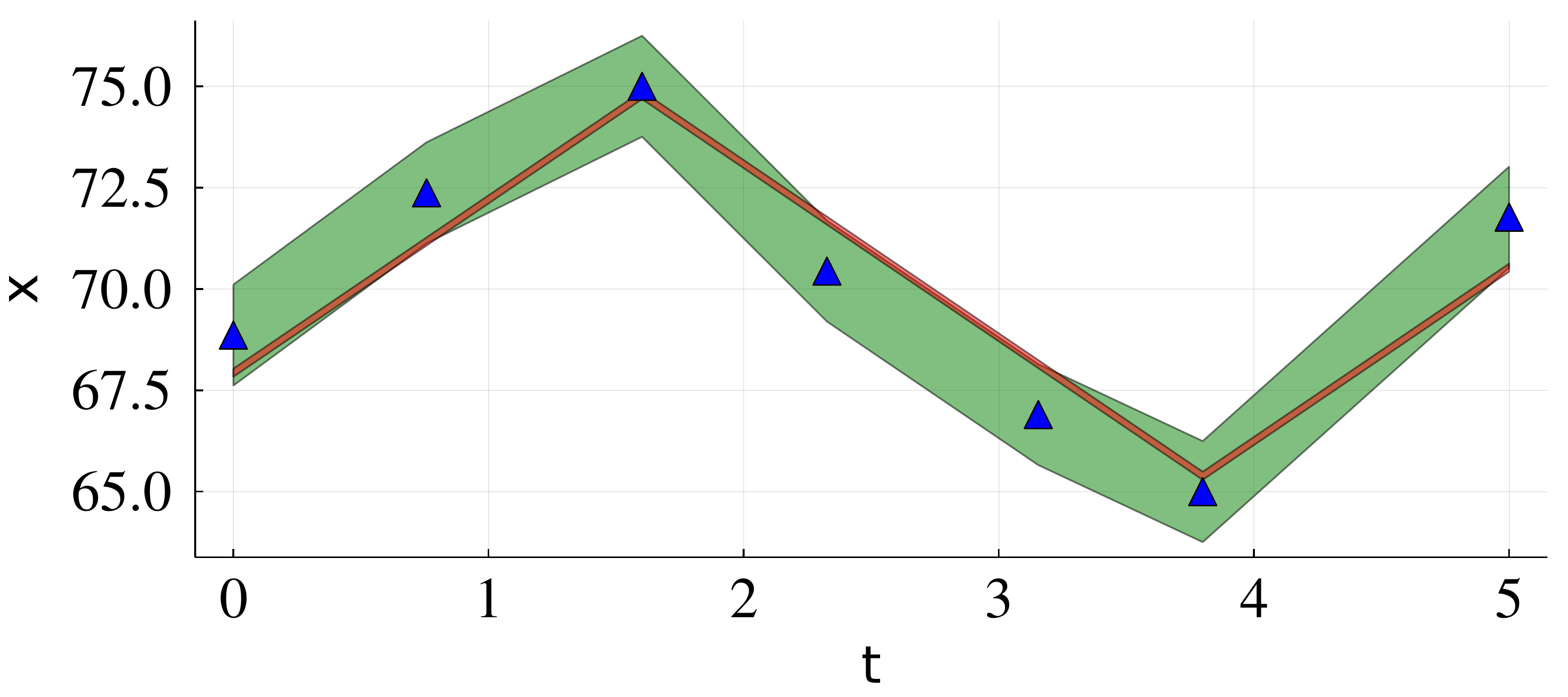}
	\hfill
	\raisebox{4mm}{\scalebox{.9}{\begin{tikzpicture}
	\node[loc,fill=\Con] (on) {\begin{tabular}{@{} c @{}}\textit{ON} \\ $\dot{x} = 4.31$ \\ $x \in [63.76, 76.24]$\end{tabular}};
	\node[loc,fill=\Coff,right=10mm of on] (off) {\begin{tabular}{@{} c @{}}\textit{OFF} \\ $\dot{x} = -4.27$ \\ $x \in [63.76, 76.24]$\end{tabular}};
	\draw[t,bend left] ($(on.east) + (0,4mm)$) to node[above] {$x \in [73.46, 76.24]$} ($(off.west) + (0,4mm)$);
	\draw[t,loop above,looseness=3] (on) to (on);
	\draw[t,bend left] ($(off.west) + (0,-4mm)$) to node[below] {$x \in [63.76, 66.24]$} ($(on.east) + (0,-4mm)$);
	\draw[t,loop above,looseness=3] (off) to (off);
\end{tikzpicture}}}
	\caption{Left: The first time series (triangle markers) inside an \eps-tube (green) and the corresponding induced execution (red), for two locations.
	Right: The synthesized \lha.}
	\label{fig:thermostat_synthesized}
\end{figure}
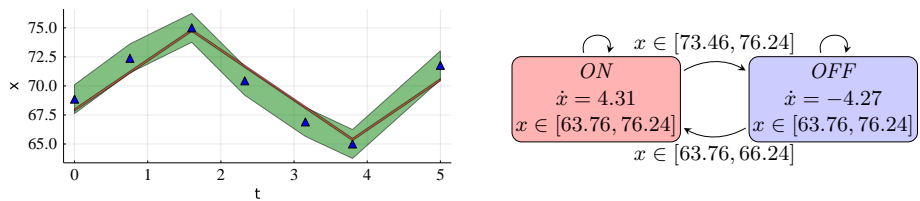

\begin{example}[cont'd]
	We intersect the two flow polyhedra and minimize the resulting polyhedron in the dimension of \eps to receive the following point:
	$m_1 = 4.31, m_2 = -4.27, x_0^{(1)} = 67.90, x_0^{(2)} = 67.63, \eps = 1.24$.
	Thus we have synthesized the following execution for the first time 
	series:
	$(\loc_1, 67.90)
	\dwell{0.76} (\loc_1, 71.18)
	\jump (\loc_1, 71.18)
	\dwell{0.84} (\loc_1, 74.80) 
	\jump{} (\loc_2, 74.80)
	\dwell{0.72} (\loc_2, 71.72)
	\jump (\loc_2, 71.72)
	\dwell{0.83} (\loc_2, 68.18) 
	\jump{} (\loc_2, 68.18)
	\dwell{0.64} (\loc_2, 65.44)
	\jump (\loc_2, 65.44)
	\dwell{1.21} (\loc_1, 70.66)$.
	The execution and the final \lha are depicted in 
	\figref{fig:thermostat_synthesized}.
	\qedex
\end{example}

\subsection{Correctness}

We show that the algorithm produces an \lha that \eps-captures the given data.

\begin{lemma}\label{lem:execution_captures}
	For every time series $s$ that is input to Algorithm~\ref{alg:synthesis}, the induced execution \eps-captures $s$, where \eps is obtained in Line~\ref{line:minimize}.
\end{lemma}

\begin{proof}
	The constraints of the flow polyhedron $\PP_s$ corresponding to $s$ enforce that the induced execution is \eps-close to all data points of $s$.
	This even holds for \emph{any} point in $\PP_s$.
	Since the concrete choice of the point in Line~\ref{line:minimize} is taken from $\PP_\H$, which is a subset of $\PP_s$, the claim follows.
\end{proof}

\begin{theorem}
	The \lha \H synthesized in Algorithm~\ref{alg:synthesis} 
	\eps-captures all time series, where \eps is obtained in 
	Line~\ref{line:minimize}.
	Furthermore, Algorithm~\ref{alg:synthesis} solves Problem~\ref{prob:synthesis_minimal} in polynomial time.
\end{theorem}

\begin{proof}
	Lemma~\ref{lem:execution_captures} ensures that the induced executions \eps-capture the time series.
	It remains to show that these induced executions belong to \H.
	This holds by construction of \H; we only sketch the main arguments.
	Each execution follows the slopes of the associated locations.
	For each location switch there exists a transition in \H.
	The executions always stay in \eps-proximity to the data points, and hence they stay inside the invariants at all times.
	Similarly, since the executions change the location at time points of the data, the guards are satisfied.
	The solution to Problem~\ref{prob:synthesis_minimal} follows from the minimization of \eps in Line~\ref{line:minimize}. For the polynomial complexity, observe that the flow polyhedron's size is polynomial in the input and that the minimization can be implemented polynomially~\cite{Khachiyan79}.
\end{proof}

We remark that the number of locations $\lambda$ and the sequence
of locations obtained from Algorithm~\ref{alg:preprocessing}
influence the quality of the \lha resp.\ the size of \eps but not the
validity of the theorem (correctness of Algorithm~\ref{alg:synthesis}).
If these inputs are unsuitably chosen, the algorithm just returns a larger value for \eps.

\begin{figure}[tb]
	\centering
	\includegraphics[width=.49\textwidth,keepaspectratio]{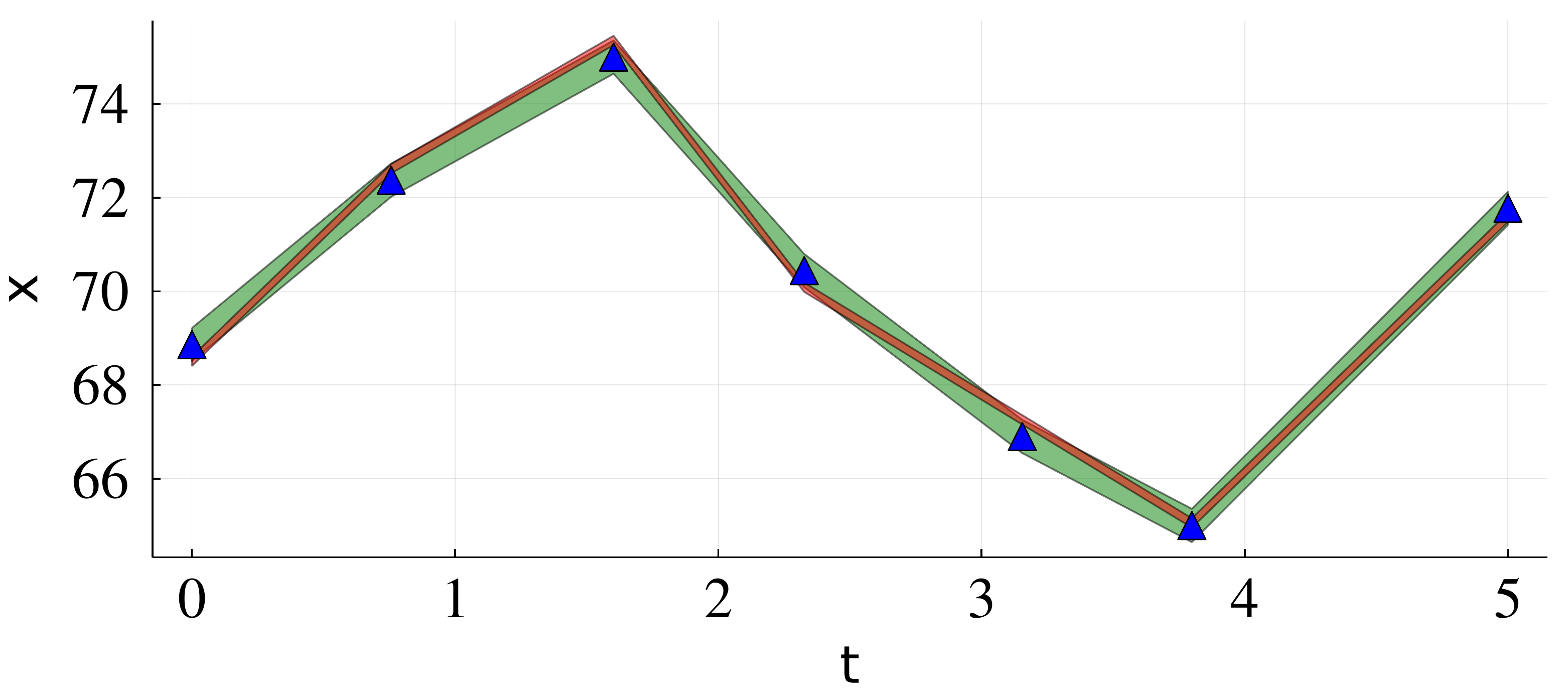}
	\hfill
	\includegraphics[width=.49\textwidth,keepaspectratio]{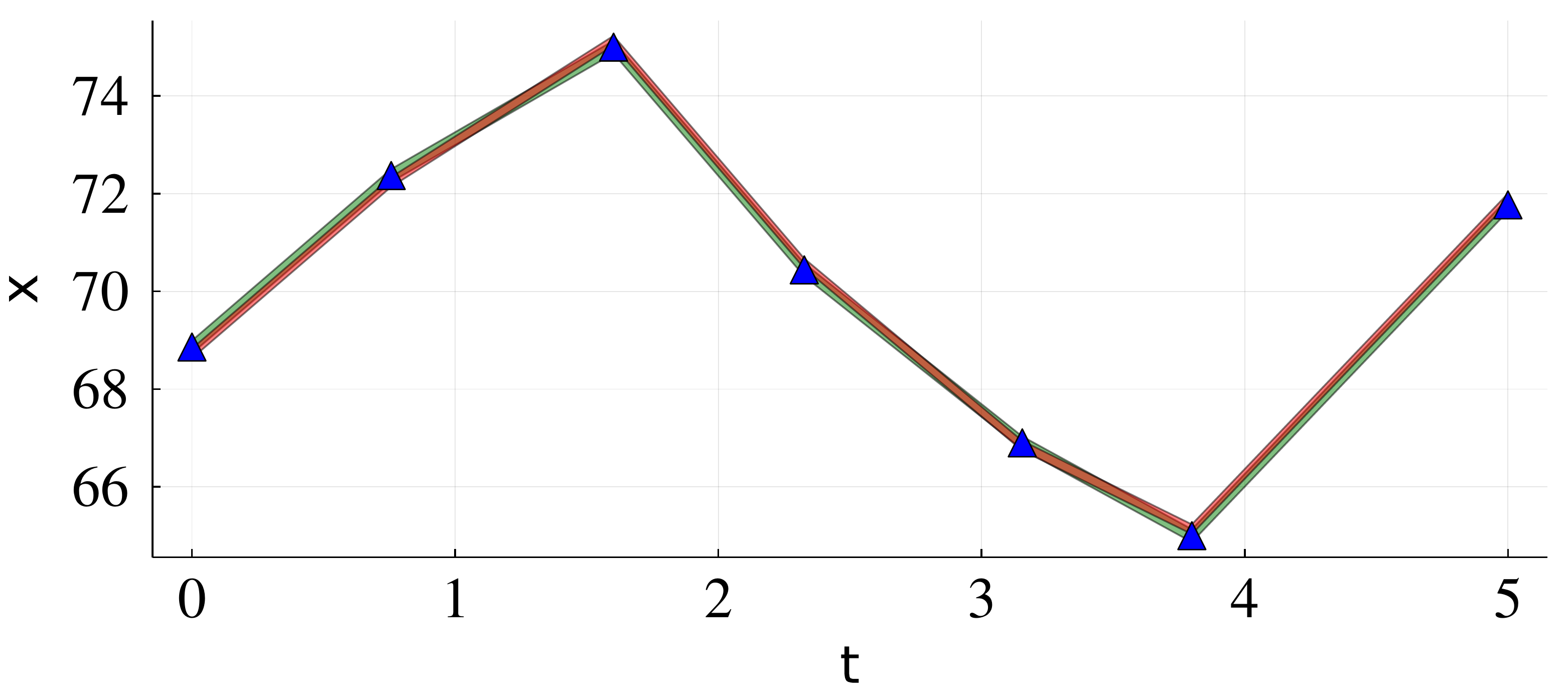}
	\caption{The first time series (triangle markers) from \figref{fig:thermostat_synthesized} inside other \eps-tubes (green) and the corresponding induced executions (red).
	Left: The result obtained for four locations ($\eps = 0.38$).
	Right: The result obtained for six locations ($\eps = 0.15$).}
	\label{fig:thermostat_synthesized_more_locations}
\end{figure}

\begin{example}[cont'd]
	\figref{fig:thermostat_synthesized_more_locations} shows the synthesized executions and corresponding values of \eps for the first time series with $\lambda = 4$ and $\lambda = 6$ locations.
	\qedex
\end{example}

\section{Evaluation}\label{sec:evaluation}

In this section we describe our implementation and present experimental results.
Our implementation in the Julia programming language is available at \url{https://github.com/HySynth/HySynthParametric}.
To generate time series, we implemented a simulator of hybrid automata based on the ODE toolbox Differential\-Equations.jl~\cite{DifferentialEquations.jl}.
For polyhedral computations we use LazySets.jl~\cite{LazySets.jl}.

\medskip

We evaluate our algorithm in two case studies.
In the first case study we investigate the scalability.
In the second case study we synthesize an LHA model on data obtained from a model of a biological system.
We note that all experiments are fully automatic with no human involved in the annotation or modeling.

\subsubsection{Scalability.}

In the first case study we measure the scalability of the algorithm in four different input dimensions: the data dimension $n$, the number of time series $r$, the number of data points per time series $p$, and the number of locations in the final automaton $\lambda$.
Here we do not use the preprocessing from Section~\ref{sec:preprocessing} for better comparability between different runs.
The majority ($> 90\%$) of the run time is spent in solving the linear program (Line~\ref{line:minimize} in Algorithm~\ref{alg:synthesis}).

To obtain the time series, we instantiate a parametric version of the thermostat model (our running example) with $n$ independent thermostats running in parallel.
We obtain $r$ random simulations of time duration $T = 40$, which are represented as time series, and then choose the first $p$ data points from them.
Since we fix $\lambda$, we pass it to Algorithm~\ref{alg:preprocessing}, which then skips the refinement procedure for $k$-means clustering in Line~\ref{line:clustering} and directly uses $\lambda$ clusters.

\begin{figure}[tb]
	\includegraphics[width=.49\textwidth]{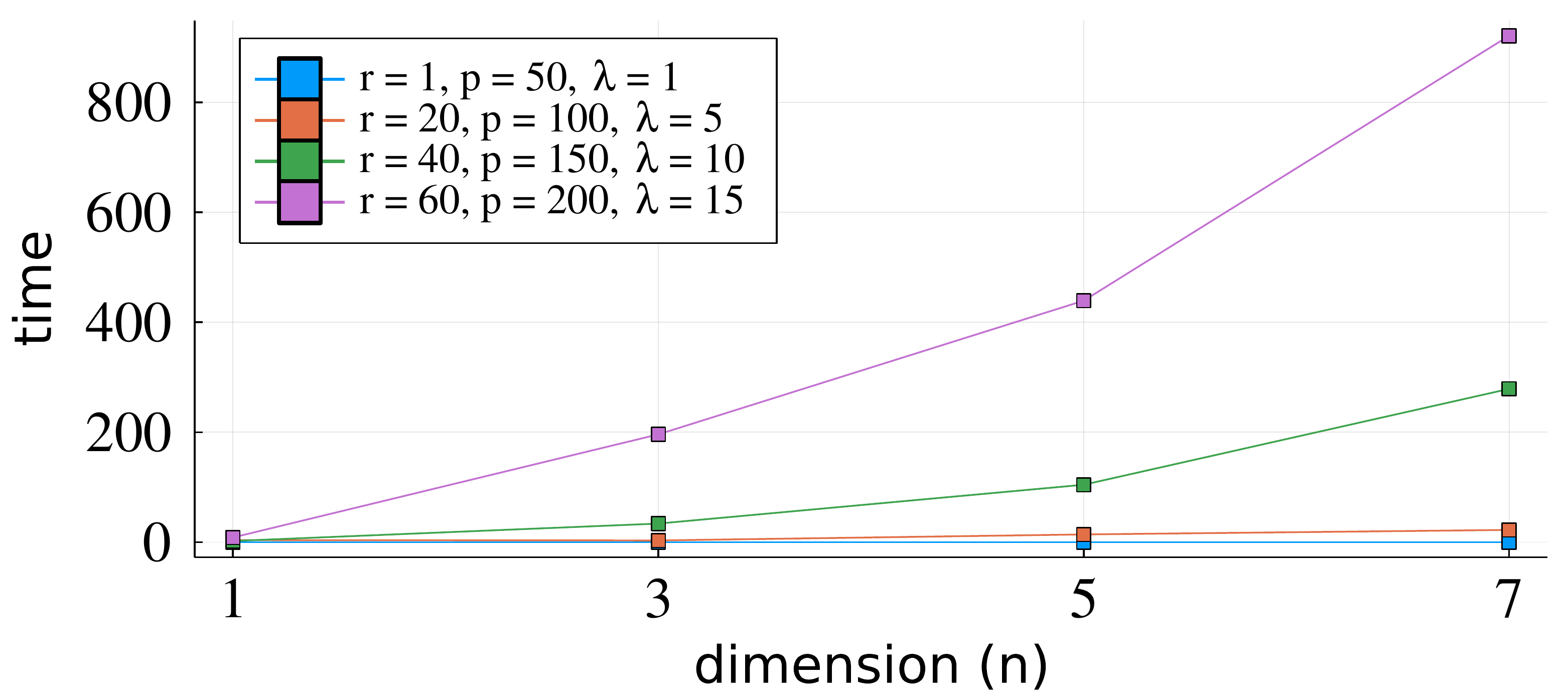}
	\hfill
	\includegraphics[width=.49\textwidth]{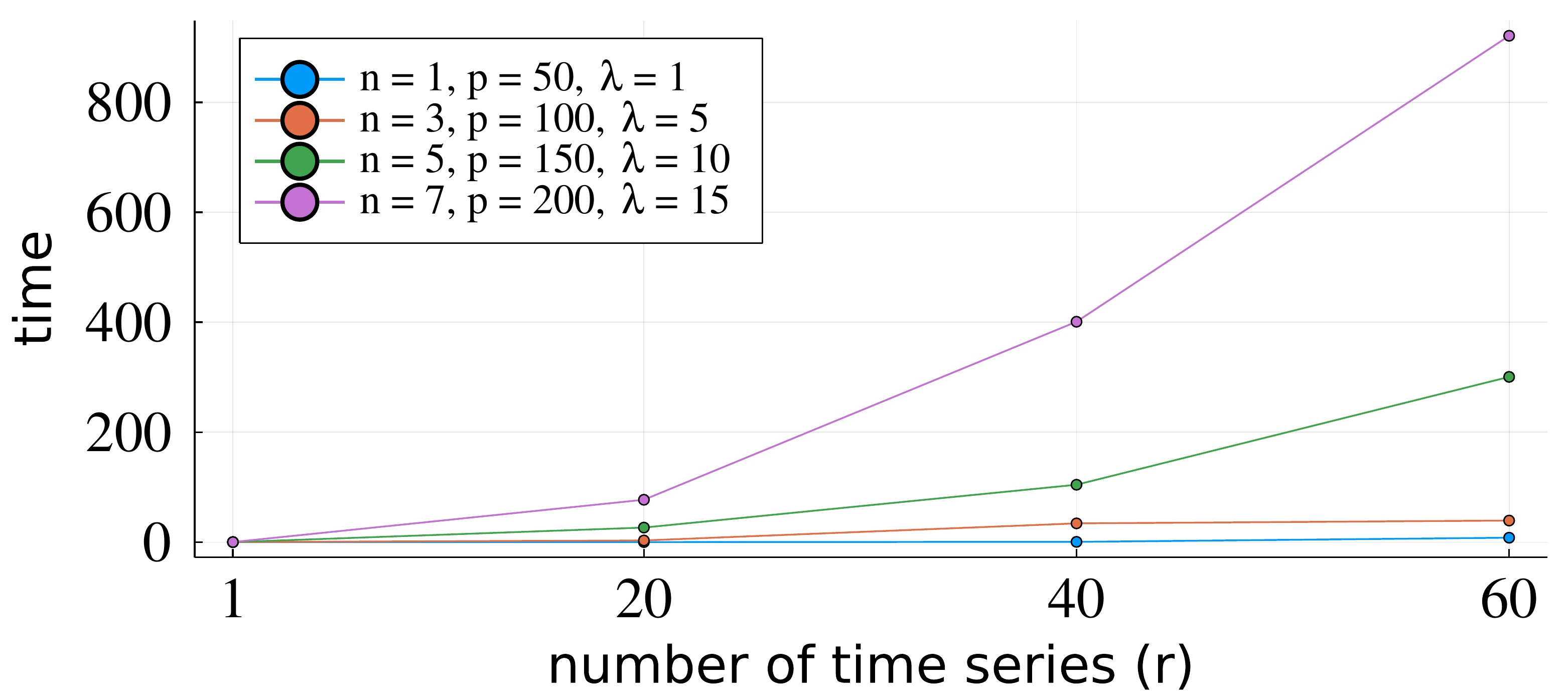}
	\\[\smallskipamount]
	\includegraphics[width=.49\textwidth]{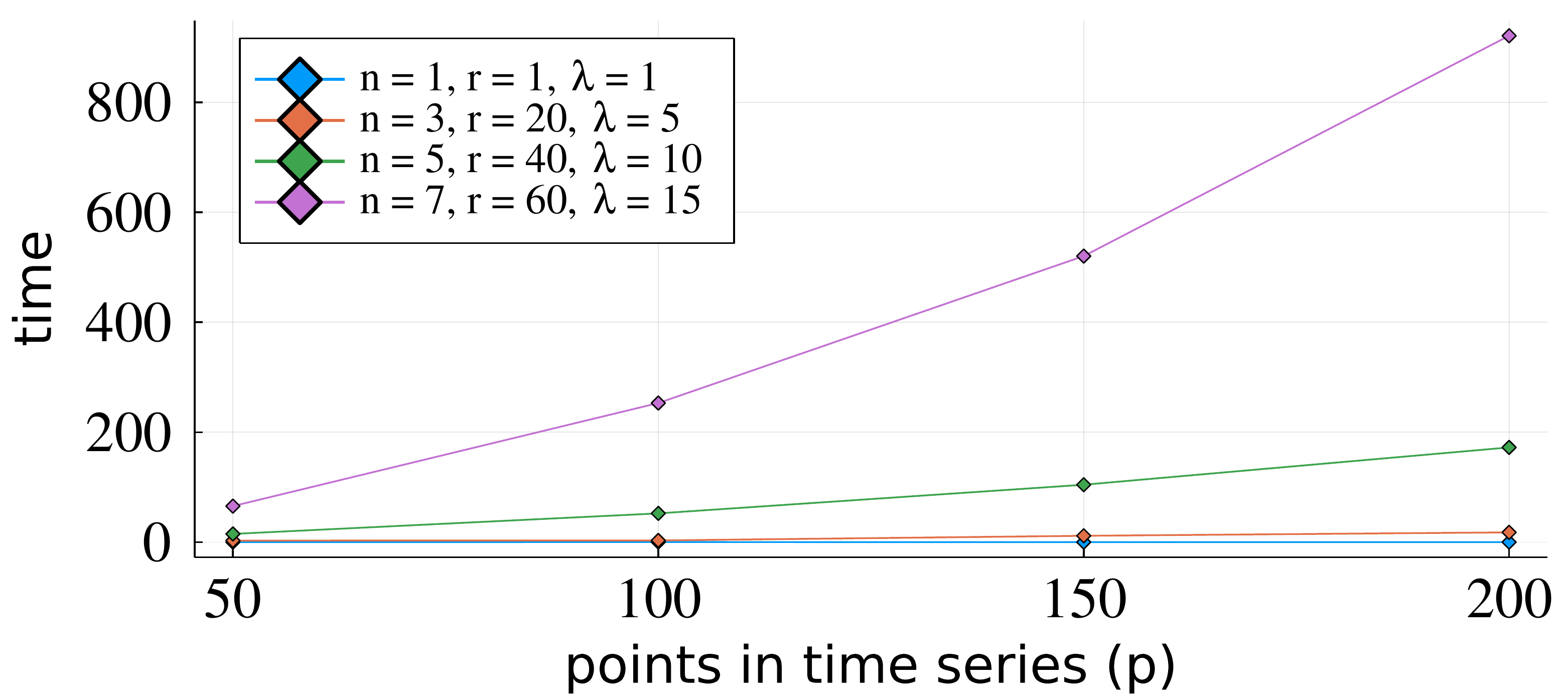}
	\hfill
	\includegraphics[width=.49\textwidth]{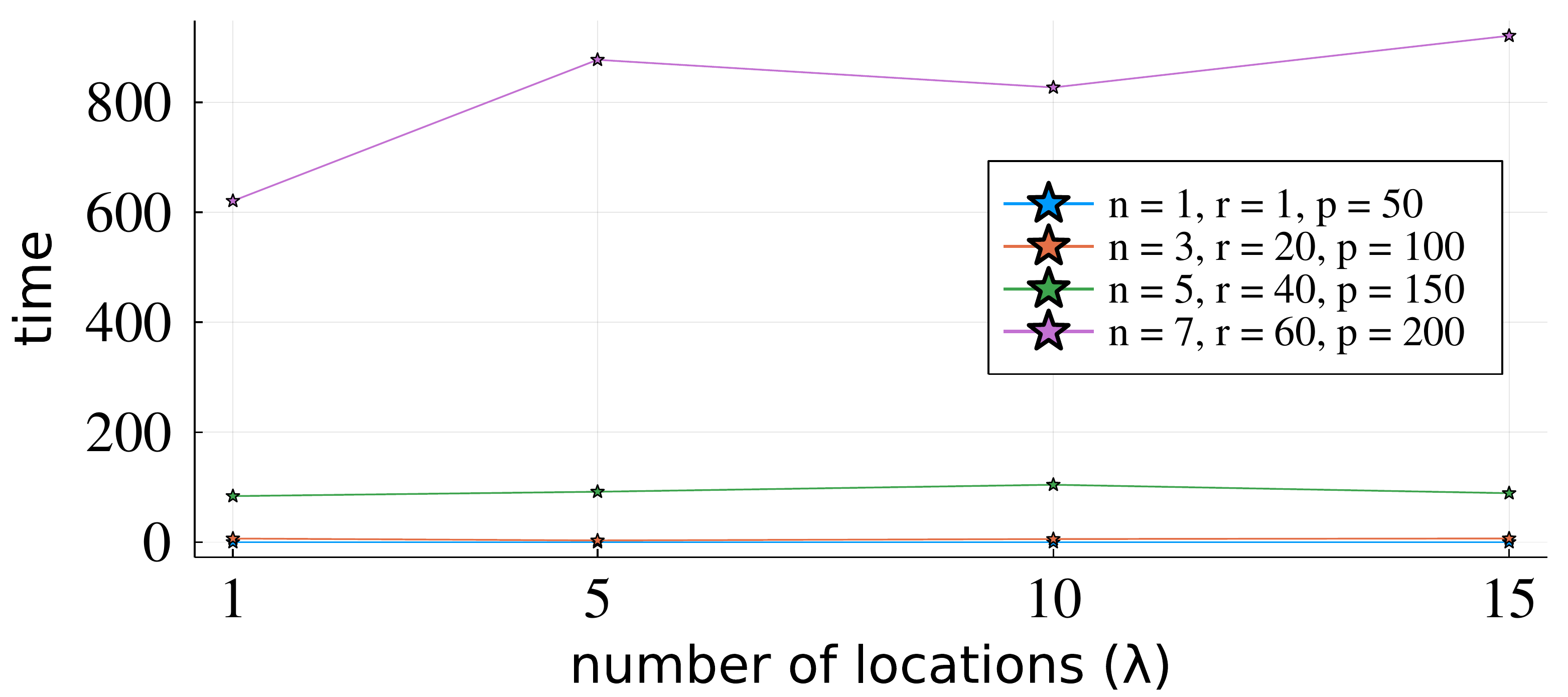}
	\caption{Scalability in four different algorithm parameters.
	Each parameter varies between four values.
	In each of the four plots we vary one parameter and fix the remaining three.
	Each plot shows four graphs with indices $i = 1, \dots, 4$, where for graph $i$ we fix the parameters to their $i$-th value (which are also given in the legend).}
	\label{fig:scalability}
\end{figure}

We consider the following combination of parameters:
$n \in \{1, 3, 5, 7\}$, $r \in \{1, 20, 40, 60\}$, $p \in \{50, 100, 150, 200\}$, and $\lambda \in \{1, 5, 10, 15\}$.
To examine the scalability in these four dimensions, we fix three parameters and plot the run time for varying only one of the parameters in Figure~\ref{fig:scalability}.

\smallskip

From the results we observe that the input parameters $n$ and $r$ have the main influence on the complexity of the problem (the corresponding graphs have the steepest growth).
The parameter $p$ is less influential, and the parameter $\lambda$ has almost no influence (the corresponding graphs barely grow and are not even monotonic).
While $\lambda$ influences the dimension of the flow polyhedron \PP, the different constraints are weakly coupled in these additional dimensions and thus the linear program is not substantially harder to solve.

In practice, when the data comes from experiments, the problem dimension $n$ is fixed, and so is $p$ if the data points are obtained from periodic measurements of fixed duration.
Increasing $r$ corresponds to additional experimental runs.
The parameter $\lambda$ can be freely chosen, but since a major benefit of hybrid automata is that they are interpretable models, we argue that $\lambda$ should not be too large.
Hence we believe that the algorithm is efficient enough to be used for real applications.
We substantiate this claim in the next case study.

\subsubsection{Regulation of a cell cycle.}

We consider the hybrid-automaton model of the regulation of a mammalian cell cycle from~\cite{SinghaniaSJT11}.
The cell cycle is modeled in nine phases.
The model has one location for each phase, affine differential equations ($\dot{\x} = A\x + \vb$), and assignments associated with some transitions.
There are three main dimensions (CycA, CycB, and CycE), one secondary dimension for the mass of the cell, and time as auxiliary dimension for time-triggered transitions.

\begin{figure}[tb]
	\centering
	\includegraphics[width=.49\textwidth]{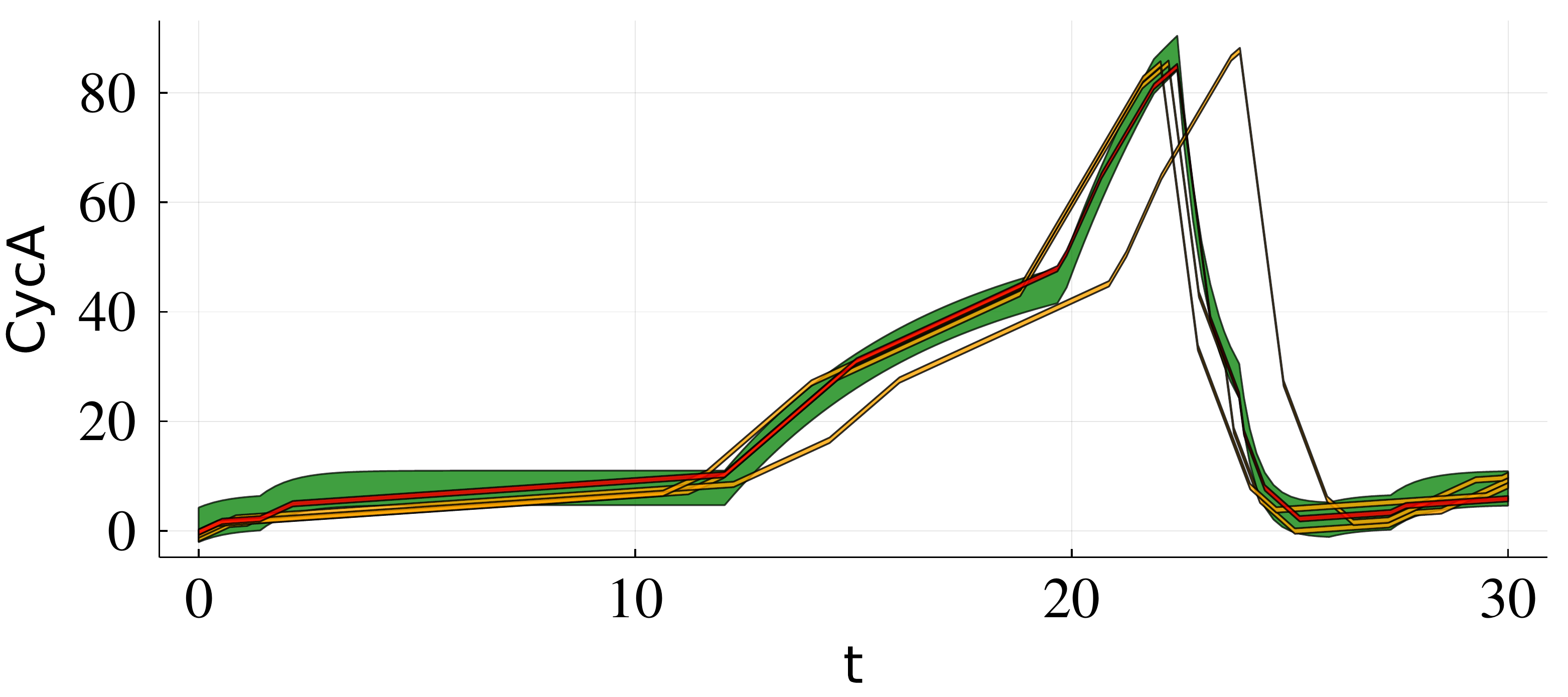}
	\hfill
	\includegraphics[width=.49\textwidth]{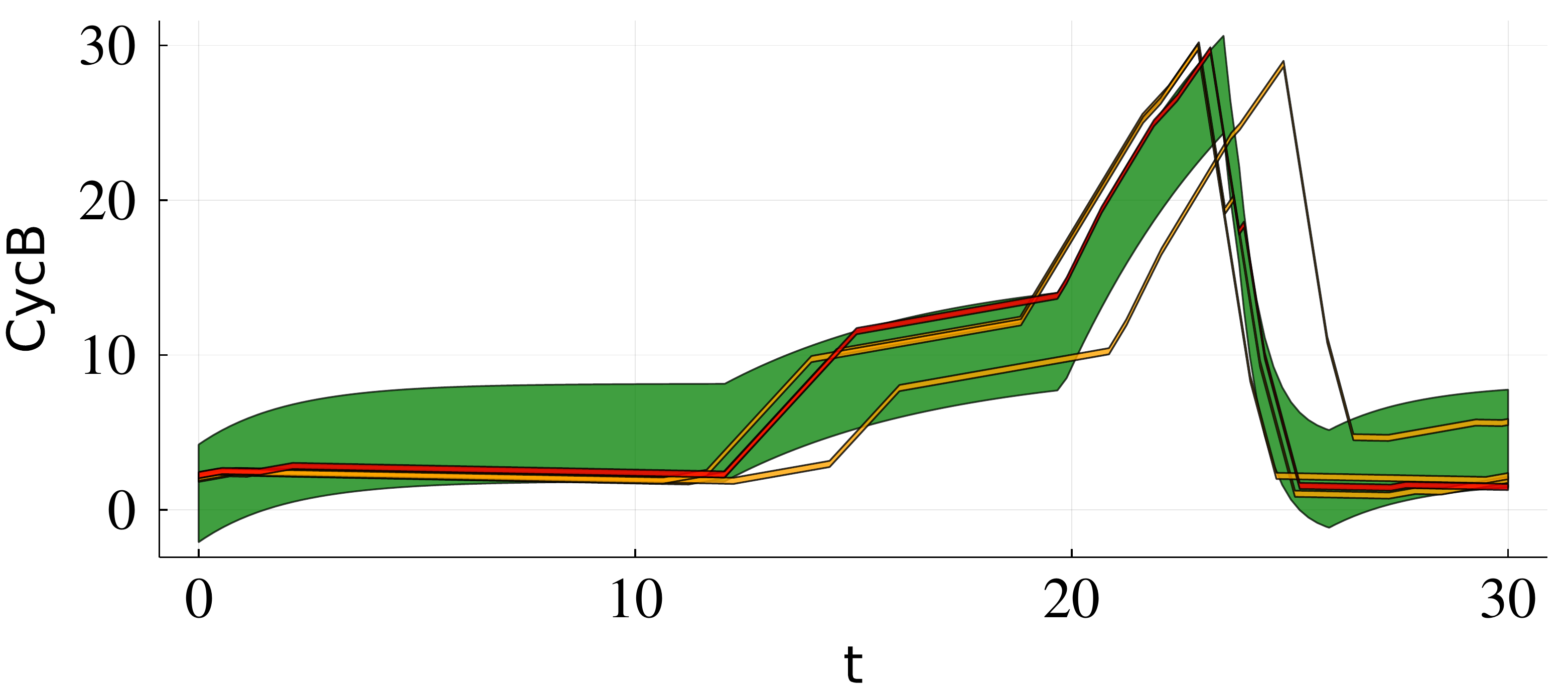}
	\caption{The first two variables of the cell-cycle regulation with the \eps-tube induced by the first time series ($\eps = 3.15$, green), the corresponding induced execution (red), and three random simulations of the synthesized model (orange).}
	\label{fig:cell_cycle}
\end{figure}

We run our synthesis algorithm on $20$ time series obtained from random simulations of the model proposed in~\cite{SinghaniaSJT11}.
In total these time series consist of $3{,}557$ data points.
Before passing them to the algorithm, we project out the time variable.
Hence our model cannot reason about time-dependent behavior.
We used the refinement process for choosing the number of locations ($\lambda$) automatically.
After $26$ seconds we obtain an LHA with nine locations and a precision value $\eps = 3.15$.
In Figure~\ref{fig:cell_cycle} we show the $\eps$-tube around the first time series together with three random simulations of the synthesized LHA.
The $\eps$-tube looks reasonably tight for the CycA dimension but wider for the CycB dimension, which is because the value of \eps is the same in all dimensions, but the plot scales differ.

\section{Conclusion}\label{sec:conclusion}

We have presented a synthesis algorithm to obtain a linear hybrid automaton from a set of time series.
The algorithm uses two independent phases.
In the first phase it constructs the discrete structure of the automaton.
In the second phase it constructs the parameter space of all possible solutions and then selects an automaton by solving a linear program.
The automaton is guaranteed to contain executions that are \eps-close to the time series, where \eps is minimal for the discrete structure chosen in the first phase.
The algorithm is polynomial and scales to thousands of data points, but it also works with scarce data.

\smallskip

We see several directions for future work.
The choice of the discrete structure in the first phase is important.
We have proposed a heuristic implementation based on clustering that does not take the number of transitions into account.
Reducing that number can remove unwanted behavior in the resulting model.

By minimizing \eps we only minimize the maximum deviation of the executions from the data points.
One can encourage the solver to find executions that stay close to the data points (the middle of the \eps-tube in the plots).
This can be encoded in the linear program by associating a cost to the sum of the deviation.

A more challenging extension is to use other classes of dynamics such as affine differential equations.
The (exponential) solutions for such systems still have a closed form.
Thus, instead of a linear program, we can solve a general optimization problem as in \cite{SotoHS21}.
The difficult part is how to select the appropriate symbolic dynamics for the different parts of the time series.

Finally, in this paper we have only considered the automatic aspects of the algorithm.
However, we believe that truly useful modeling ultimately requires 
interaction with a human in the loop.
The separation of concerns -- first finding a suitable discrete structure and formulating a parametric solution for finding suitable continuous dynamics -- allows scientists to incorporate domain knowledge, e.g., by adding further modeling constraints beyond \eps-capturing.
A key question is how to refine the model if the results are not accepted.

\subsection*{Acknowledgements} This work was supported in part by the European Union’s Horizon 2020 research and innovation programme under the Marie Skłodowska-Curie grant agreement no.\ 847635, by the ERC-2020-AdG 101020093, by DIREC - Digital Research Centre Denmark, and by the Villum Investigator Grant S4OS.

\bibliographystyle{splncs04}
\bibliography{bibliography_arxiv}

\end{document}